\documentclass[a4paper]{article}
\usepackage{IS2020_paper_kit/LaTeX/INTERSPEECH2020}
\usepackage{amsmath}
\usepackage{graphicx}
\usepackage{amsfonts}
\usepackage{comment}
\usepackage{multirow}
\usepackage{hyperref}
\usepackage{blindtext}
\usepackage{blindtext,letltxmacro,xcolor,xparse}

\title{Jointly Trained Transformers models for Spoken Language Translation}
\name{Hari Krishna Vydana, Martin Karafi\'{a}t, Katerina Zmolikova, Luk\'{a}\v{s} Burget, ``Honza'' \v{C}ernock\'{y} \thanks{The work was supported by Czech National Science Foundation (GACR) project ``NEUREM3 No. 19-26934X}}
\address{\mbox{\hspace{-0.2cm}Brno University of Technology, Faculty of Information Technology, IT4I Centre of Excellence, Czechia}}
\email{vydana@fit.vutbr.cz}

\LetLtxMacro{\blindtextblindtext}{\blindtext}
\LetLtxMacro{\blindtextBlindtext}{\Blindtext}

\RenewDocumentCommand{\blindtext}{O{\value{blindtext}}}{%
  \begingroup\color{gray}\blindtextblindtext[#1]\endgroup
}
\RenewDocumentCommand{\Blindtext}{O{\value{blindtext}}O{\value{Blindtext}}}{%
  \begingroup\color{gray}\blindtextBlindtext[#1][#2]\endgroup
}

\begin{document}
%
\maketitle
\begin{abstract}
Conventional spoken language translation~(SLT) systems are pipeline based systems, where we have an Automatic Speech Recognition~(ASR) system to convert the modality of source from speech to text and a Machine Translation~(MT) systems to translate source text to text in target language. Recent progress in the sequence-sequence architectures have reduced the performance gap between the pipeline based SLT systems (cascaded ASR-MT) and End-to-End approaches. Though End-to-End and cascaded ASR-MT systems are reaching to the comparable levels of performances, we can see a large performance gap using the ASR hypothesis and oracle text w.r.t MT models. This performance gap indicates that the MT systems are prone to large performance degradation due to noisy ASR hypothesis as opposed to oracle text transcript.\\
In this work this degradation in the performance is reduced by creating an end to-end differentiable pipeline between the ASR and MT systems. In this work, we train SLT systems with ASR objective as an auxiliary loss and both the networks are connected through the neural hidden representations. This training would have an End-to-End differentiable path w.r.t to the final objective function as well as utilize the ASR objective for better performance of the SLT systems. This architecture has improved from BLEU from 36.8 to 44.5. Due to the Multi-task training the model also generates the ASR hypothesis which are used by a pre-trained MT model. Combining the proposed systems with the MT model has increased the BLEU score by 1. All the experiments are reported on English-Portuguese speech translation task using How2 corpus. The final BLEU score is on-par with the best speech translation system on How2 dataset with no additional training data and language model and much less parameters. 
\end{abstract}

\section{Introduction}
\label{sec:intro}
Spoken Language Translation (SLT) refers to the task of transcribing the spoken utterance in source language with the text in the target language. SLT systems are typically categorized in two types: Cascade systems and End-to-End systems. Cascade SLT systems in their popular form comprises of an automatic speech recognition~(ASR) system followed by a machine translation~(MT) system. The initial ASR system generates the text sequence for the spoken utterance and the generated text sequence is translated to the target language by a machine translation~(MT) system. Speech recognition task has a monotonic alignment between the spoken sequence and the labeling sequence. But this property is not present in MT and SLT tasks. ASR system relies more on the local context to transcribe the speech while MT systems would need the wider context of a word to translate it. In a cascade approach, both ASR and MT could be trained separately. The improvements in both ASR and MT performance could be easily translated to the SLT model. End-to-End SLT system is a single model that directly transcribes the spoken utterance with the text in target language. End-to-End SLT models have to learn the complex mapping between the spoken sequence and the labeling sequence in the target language which involves the word splitting, merging and reordering. End-to-End SLT systems could result in simpler models to train, have low latency and could lead to direct optimization of the required task. Using the the source language text while optimizing the End-to-End SLT model could improve the performance.\\

Presently cascade~SLT systems perform better than the End-to-End SLT systems. Popular End-to-End SLT systems are sequence-to-sequence with attention models. With the recent advancements in these models the performance gap between cascade and End-to-End SLT systems is decreasing. But the cascade SLT system suffers from a large performance gap when MT models are evaluated using ASR hypotheses compared to using the oracle text. This large performance gap is due to the erroneous ASR hypotheses and the MT system which is not able to handle the errors ASR hypotheses. In this study, we explore approaches to reduce this performance gap by creating an End-to-End differentiable pipeline between ASR and MT systems. SLT systems are trained with ASR objective as an auxiliary loss and both the networks are connected through the neural hidden representations. By this training the model could have differentiable path between input(speech) and the final labeling sequence and also utilize the ASR objective for better performance of SLT system. During the inference, both the models are connected through N-best hypotheses i.e., N-best hypotheses from ASR model are translated by the MT system in parallel. The likelihood of every N-best translation is conditioned on the likelihood from the ASR hypothesis as mentioned in equation~\ref{coupling_decoders}.

Presently most state-of-the-art MT systems use transformer model~\cite{Transformer}. Transformer is a recurrent-free encoder-decoder model. Transformer employs self-attention to model the dependencies in the sequences and cross-attention to model the co-relations across the sequences. With the use of causality masks to prevent the model from peaking in to the future, transformer model could be parallelized while training. The model has shown capabilities to model the long-term contexts. Transformer model has been adapted for training ASR systems~\cite{dong2018_speech_transformer,ESPNET_tranformer}. Transformer models have been used for developing End-to-End SLT systems~\cite{SPMT_tranformer_diag_loss,ESPNET_rnn_tranformer}. In this study, we use transformer models for training ASR, MT and and SLT systems. The performances of End-to-End and cascade SLT models are analyzed. The SLT systems in this study are developed with two different granularities i.e., character/BPE units. The performances of these models are analyzed on English to Portuguese translation tasks. 

\paragraph*{Recent Developments in SLT systems}
Some of the recent SLT models submitted to IWSLT-2019 are described below:\\
The cascade SLT system with a pipeline of ASR, punctuation model and MT has the best performance in IWSLT 2019~\cite{KIT_IWSLT2019}.
The ASR used in this work is an ensemble of LSTM-based encoder decoder model and a big transformer model(>150M parameters) trained with stochastic depth~\cite{Transformer_stocastic_depth}. The MT model used is a multi-lingual model(>200M) trained for two languages(German and Portuguese). End-to-End SLT systems are trained using LSTM-encoder-decoder models with characters as target units in~\cite{ON_TRAC_IWSLT2019}. Transformer models have been adapted for training End-to-End SLT systems~\cite{di2019adapting}. A convolutional front-end is used to sub-sample the acoustic sequence by a factor of 4 and the models are optimized with an addition distance penalty for the attention to focus more on the local context. Augmentation methods such as spec-augment~\cite{Spec_Aug} and use of back-translated synthetic text has improved the performance~\cite{FBK_iwslt2019}. The performances  End-to-End SLT systems have been improved by initializing the parameters of the model from independently trained ASR and MT systems~\cite{inaguma2018jhu}. Block multi-task learning~(BMTL) has been used for training MT systems~\cite{CMU_IWSLT2019}. BMTL is an encoder-decoder networks where encoder is shared by decoders which generates tokens with multiple granularities such as characters,sub-words. A Multi-Engine Machine Translation~(MEMT) is a machine translation that uses inputs in multiple granularities and produces word based translations. Use of pre-trained language models in encoder decoder models have been explored for developing MT models in~\cite{LIG_IWSLT2019_using_bert}. Multi-model SLT systems were studied in ~\cite{Multimodel_MT_IMP_LONDON}, an LSTM-based encoder-decoder ASR is followed by a multi-model Transformer model. Text and video features have been used for developing multi-model machine translation systems.

\section{Database}
\label{Database}
All the experiments in this paper are conduced using HOW2 data-set~\cite{how2_dataset_paper}. The data-set comprises of train, dev and test sets. The train, dev and test sets consists of 185K, 2305 and 2022 sentences which amounts to 298, 3 and 4 hours of speech data. All the sentences in the data-set have parallel Portuguese translations. All the models in this work are trained using train set and early stopping is done using validation set. 
\paragraph*{Pre-processing and Feature Extraction:}
All the text is lower-cased and the punctuation symbols are removed. The sentence piece model is trained to obtain a sub-word vocabulary of 5000 tokens for both English and Portuguese texts. Audio from the videos are extracted and 16~kHz signal is used in this experiments. 40-Dimensional Mel-filter bank features are extracted using Kaldi-toolkit with 25~ms window size and 10~ms overlap. Cepstral mean and variances of the features are normalized per-video. The performances of all the translation systems are measured using Sacre-BLEU.   
\section{Transformer Model}
\label{Transformer_Model_description}
ASR systems used in this work are trained using transformer models. Transformer is an encoder-decoder model that models the temporal dependencies using self-attention mechanism. Typical encoder-decoder model has three blocks i.e., encoder, decoder and an attention module. The encoder models the sequential dependencies in the input and produces a hidden state sequence H. The decoder in conditioned on the past predicted label sequence $y_1, y_2 ,y_3 ,..., y_i$ and the context vector $c_i$ to produce the present label $y_{i+1}$. The context vector $c_i$ is produced by summarizing the sub-sequence H which is relevant to produce the correct label. Attention mechanism provides a way to peek in to the input sequence and the predicted output sequence while generating a new label.   

Let $X$=[$x_1$, $x_2$, $x_3$ ,......., $x_T$] be the input acoustic features and H=[$h_1$, $h_2$, $h_3$, ........, $h_T$] be the hidden representation. The hidden representation H is used by the decoder to produce the label sequence Y=[$y_1$, $y_2$, $y_3$, ......, $y_L$]. 

Self-attention mechanism produces the output sequence as a weighted sum of the input sequence. Where the weights are derived from the similarity between each query and key vectors. The self-Attention mechanism can be described by the below equation:\\
\begin{equation}
\label{QKV_equation}
\mathrm{Attention(Q,K,V)}=\mathrm{softmax(\frac{QK^{T}}{\sqrt{d_k}}) V.}
\end{equation}
Where Q,K and V are the query, key and value respectively. Here query, key has the same dimensionality and key, value has the same sequence length. The scalar $\frac{1}{\sqrt{d_k}}$ is a normalizing factor to prevent the softmax from entering to the regions with small gradients.

The Self-attention mechanism does not have an explicit notion of position in the sequence. The information about the position in the sequence is conveyed by fixed trigonometric position embeddings. The position embeddings used in this work are described below:
\begin{equation}
\label{positional_encoding_equation}
\mathrm{PE_{(pos,i)}}=\mathrm{\left\{\begin{matrix}
sin(\frac{pos}{10000^{2i/d_{model}}});\:\:\:\:\:\:\:\:\:       0 \leq i < d_{model}/2\\ 
cos(\frac{pos}{10000^{2i/d_{model}}});\:\:\: d_{model}/2 \leq i < d_{model}
\end{matrix}\right.}
\end{equation}
In the above sequence, $i$ is the index of dimension and $pos$ represents the position in the sequence. The encoding scheme generates a vector for each position and this vectors are summed to the input feature sequence. 

\begin{figure}[!ht]
  \includegraphics[width=80mm,height=80mm]{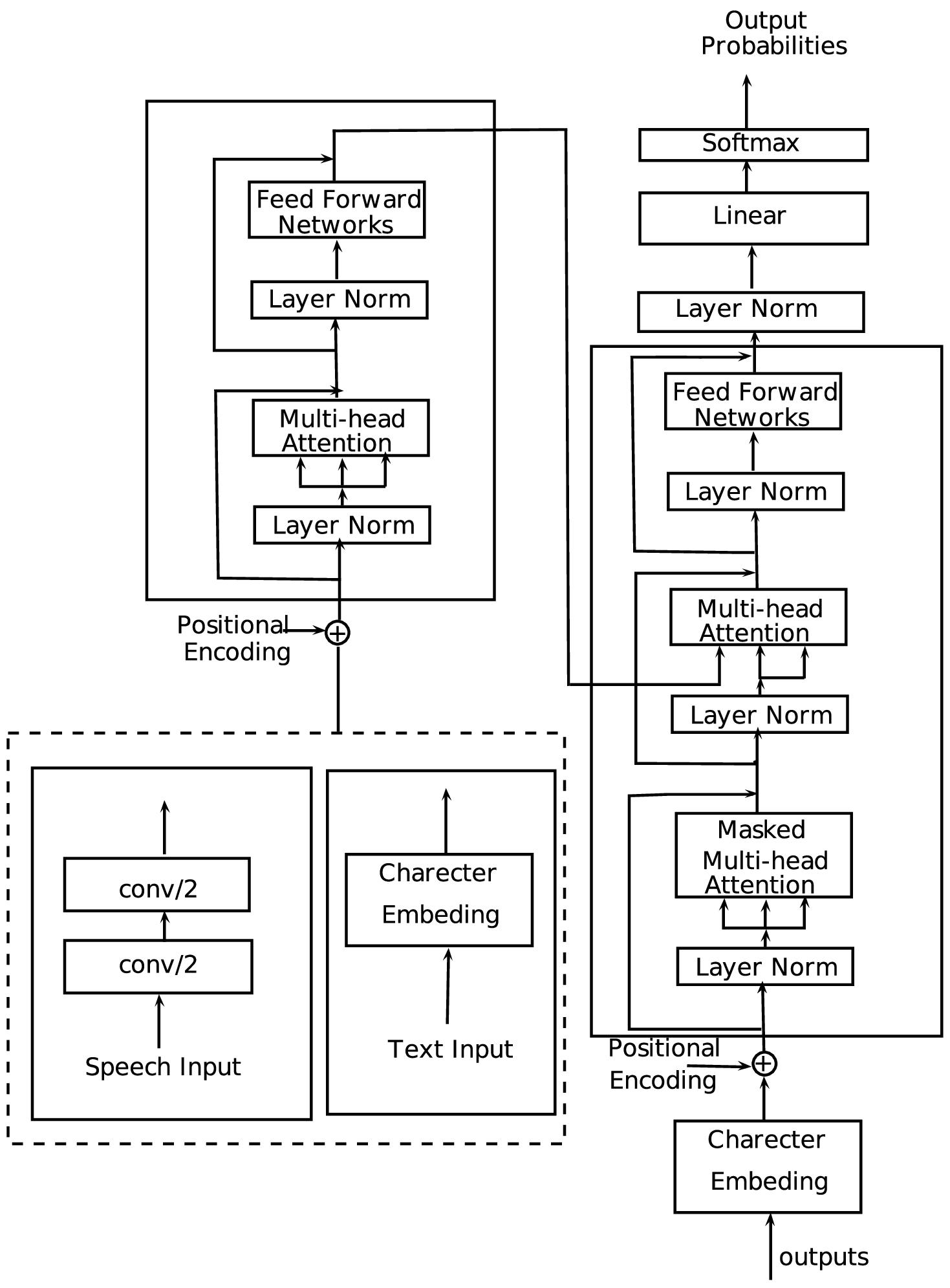}
\caption{Block diagram of the Transformer models used in this study.}
\label{Transformer-block-diagram}
\end{figure}

Transformer employs a Multi-Head attention to learn the dependencies between the same sequence and different sequences. The Multi-Head attention computes scaled dot product attention multiple times in parallel. Before computing the attention, queries, keys and values are passed through linear projections. Each time the scaled dot-product attention is computed, the outputs of all the heads are concatenated and passed through a linear layer with dimension $d_{model}$. 
\begin{equation}
\label{Multi_head_attention}
\mathrm{MultiHead(Q,K,V)}=\mathrm{Concat[head_1, head_2 ,..., head_h]W^{O}}
\end{equation}
where each head is given by \\
\begin{equation}
\mathrm{h_i=Attention(QW_{l}^{q},KW_l^{k},VW_l^{v})}
\end{equation}
A position-wise feed-forward network consists of two linear layers with rectified linear activation units~(ReLU)  as shown in the equation\ref{eq_pos_ff}.
\begin{equation}
\label{eq_pos_ff}
 \mathrm{FFN(x)}=\mathrm{max(0,W_1x+b_1)W_2+b_2}
\end{equation}
Where $W_1\in\mathbb{R}^{d_{model}\times d_{ff}}$ and $W_2\in\mathbb{R}^{d_{ff}\times d_{model}}$, $b_1$,$b_2$ are the biases for the weights $W_1$ and $W_2$ respectively. Here $d_{ff}$ is the dimensionality of position-wise feed-forward network. The blocks described above are connected as shown in the Figure.~\ref{Transformer-block-diagram}.

While training with speech inputs (i.e., for ASR and End-to-End SLT the models) convolution front-end is used. The front-end convolution module contains two convolution layers with a kernel size of 3 and stride of 2. The convolution layers used have 64 output channels and are trained with ReLU. These layers reduce the sequence length of the speech features and memory requirements of the transformer. The positional encoding is added to the features after the convolution layers. While training with text inputs (i.e.,for MT systems) an embedding layer is used as the input layer to convert the discrete symbol sequence to the sequence of continuous embeddings.

The model is optimized using ADAM optimizer with $\beta_1$=0.9 and $\beta_2$=0.99 and $\epsilon = 10^{-9}$. The learning rate schedule used is defined below:
\begin{equation}
\label{learningrate_schedule}
 \mathrm{lrate=k \times d_{model} \times min(step^{-0.5},step \times warmup^{-1.5})}
\end{equation}
The models are trained with 25000 warmup steps and the value of k is set to 1. A label smoothing of 0.1 is used for training the models. Both the residual and attention dropout values are set to 0.1.

\section{Comparing the Performance of Cascade and End-to-END SLT Systems}
\subsection{Transformer ASR systems}
\label{section_Transformer_ASR}
ASR systems are trained with characters or BPE units as target units. Both the models are trained with 12 encoder layers and 6 decoder layers. The models are trained for 150 epochs. Each batch comprised of 7000 target units. The models from the last ten epochs are averaged and the averaged weights are used during the decoding. ASR hypothesis are decoded with a beam size of 10. The start-of-the-sequence(<SOS>) and end-of-the-sequence (<EOS>) are modeled by additional tokens. The data-set has some very long sequences with 400-500 characters and decoding these sentences increases the decoding time. To reduce the decoding time the vectorized beam search described in~\cite{vectorized_beam_search} has been used. Decoding with larger beam size has been counter productive, the prefix with the noisy EOS token has higher log-likelihood compared to the actual sequence in the beam. A threshold mechanism described in~\cite{selftraining_EOS_threshold_FAIR} has been used .i.e., the EOS is considered only if the probability of EOS is $\gamma$ times the top candidate. The value of $\gamma$ is set to be 1 and this is obtained by fine tuning on the dev set. The length penalties of 1 and 0.8 are used for character and BPE models. No language model has been used in this work. The performance of both the models presented below Table.~\ref{Transformer_ASR}. 

\begin{table}[!ht]
\renewcommand{\arraystretch}{1.5}
\setlength{\tabcolsep}{2pt}
\scriptsize
\begin{center}
\caption{ASR systems trained using Transformer Models.}
\label{Transformer_ASR}
\begin{tabular}{|c|c|c|}
\hline
~&\multicolumn{1}{|c|}{Dev set(WER)}&\multicolumn{1}{|c|}{Test set(WER)}\\
\hline
TDNN-LFMMI &-& 13.7\\
\hline
S2S-Attention~\cite{how2_dataset_paper}&-&19.4\\
\hline
Transformer (character) &17.47 & 17.87\\
\hline
Transformer (BPE) & 16.85 & 16.52\\
\hline
\end{tabular}
\end{center}
\end{table}

Row 3 of Table.1 is the performance of the LSTM-based sequence-to-sequence model presented in~\cite{how2_dataset_paper}. Row 4 \& 5 are the performances of ASR systems developed with transformer models with characters and BPE~(5K) units as target units. The performance of transformer model with BPE units as targets has performed better. Row 2 is the ASR systems trained with TDNN-LFMMI using Kaldi-recipes this ASR is not used in any SLT systems, it is presented to compare the difference in the performances of the ASR systems.

\subsection{Transformer Machine Translation Systems}
\label{section_Transformer_machiene_translation}
Transformer models have been used for training Machine Translation~(MT) systems. MT systems are trained with three different input-output granularities such as characters-characters, characters-BPE and BPE-BPE. The details about the architecture of the networks are described as follows: The model has 6-encoder layers, 6-decoder layers and the dimensionality of $d_{model}$ and $d_{ff}$ are 512, 1024 respectively. The models are trained with 8 parallel heads. The models are trained for 150 epochs and each batch comprised of 7000 target units. The model from the last ten epochs are averaged and the averaged weights are used for decoding. The beam size and length peanlities for decoding are derived from the development set. The optimal beam size for all the granularities is 5, but the length penalties has varied, for characters-characters, characters-BPE and BPE-BPE the length penalties are 1.2, 1, 1.2 respectively. The tokens predicted from the model are converted back to the text and the Sacre BLEU is computed. The EOS threshold described in section~\ref{section_Transformer_ASR} has been used. The performances of the MT systems trained are presented in Table.~\ref{Transformer_MT}.

\begin{table}[!ht]
\renewcommand{\arraystretch}{1.5}
\setlength{\tabcolsep}{2pt}
\scriptsize
\begin{center}
\caption{MT systems trained using Transformer Models.}
\label{Transformer_MT}
\begin{tabular}{|c|c|c|}
\hline
~&\shortstack{Dev\\Sacre-BLEU}&\shortstack{Test\\Sacre-BLEU}\\
\hline
S2S-Attention-(BPE)&~&54.4~\cite{visual_modality_for_MT}\\
\hline
Transformer-(characters-characters)&53.08&52.08\\
\hline
Transformer-(BPE-BPE)&53.16&52.01\\
\hline
Transformer-(characters-BPE)&55.32&54.80\\
\hline
\end{tabular}
\end{center}
\end{table}

From Table.\ref{Transformer_MT}, it can be observed that both character-character and BPE-BPE based models had comparable performances. The charecter based models have taken longer time to train and decode compared to the BPE models. The models with character input and BPE as output has performed better than the other two systems.

\subsection{Cascade SLT models}
Cascaded SLT comprises of an ASR system followed by an MT. Two different cascades systems have been trained i.e., ASR systems with characters, BPE's as output units along with the corresponding MT model. The performance of cascaded SLT system is presented in the Table.~\ref{Cascaded_ASR_MT_Transformer}. The ASR and MT models described in section~\ref{section_Transformer_ASR}, section~\ref{section_Transformer_machiene_translation} are used in the cascaded pipeline. The ASR hypothesis is decoded and the decoded hypothesis is used by the MT model to produce the translation.

\begin{table}[!ht]
\renewcommand{\arraystretch}{1.5}
\setlength{\tabcolsep}{2pt}
\scriptsize
\begin{center}
\caption{Cascaded Sp-MT models trained using Transformer.}
\label{Cascaded_ASR_MT_Transformer}
\begin{tabular}{|c|c|c|c|c|c|c|}
\hline
\multirow{2}{*}{\shortstack{Architecture \\ (Input-Output) Granularity}} ~&\multicolumn{2}{|c|}{1-best}&\multicolumn{2}{|c|}{n-best}&\multicolumn{2}{|c|}{ranked n-best}\\
\cline{2-7}
~&Dev set&Test set&Dev set&Test set&Dev set&Test set\\
\hline
\shortstack{Transformer\\(characters-characters)}&39.12&39.18&-&-&-&-\\
\hline
\shortstack{Transformer\\(characters-BPE)}&41.21&41.31&39.72&40.73&42.52&42.31\\
\hline
\shortstack{Transformer\\(BPE-BPE)}&41.86&41.71&39.16&39.56&43.68&43.6\\
\hline
\end{tabular}
\end{center}
\end{table}
From Table~\ref{Cascaded_ASR_MT_Transformer}, a large performance degradation can be observed between the cascade SLT systems and the MT systems described in section~\ref{section_Transformer_machiene_translation}. Due to the use of ASR hypothesis as input text the performance of MT systems a degradation in the BLEU > 10 has been observed. Column~1 is the cascade-SLT system, Column 2, 3 is the performance of SLT systems using n-best hypothesis. Columns 4, 5 are the performance of SLT systems which uses the n-best hypothesis which are ranked based on the log-likelihood scores. The per-sentence log-probabilities from the n-best ASR hypothesis is used as a ranking score. The prefixes that correspond to the ASR hypothesis are initialized with their corresponding n-best ranking score. The general pipeline search is described below:
\begin{equation}
\label{coupling_decoders}
\begin{array}{l}
P(y|x)=\arg\max_{y}\sum_{z \in \hat{Z}(x))}P(y|z)P(z|x)))\\
\equiv \arg\max_{y}\sum_{z \in \hat{Z}(x))}log(P(y|z))+log(P(z|x))))
\end{array}
\end{equation}
In the above equation, $x$ is source speech and $y$ is the target token sequence and $z$ is the source token sequence. Using the N-best hypothesis without ranking scores from the ASR has degraded the performance which can be seen in column~3,4. Using N-best hypothesis with ranking scores has improved the performance. The training procedure described in section~\ref{joint_traing} is training the above objective function by considering only the one-best hypothesis.

\subsection{End-to-End SLT models}
End-to-End SLT systems are trained using transformer models. The architectural details of the model are described below: the model has 12-encoders, 6 decoder layers with $d_{model}$, $d_{ff}$ are 256 and 2048 with 4 heads. The models are trained to predict the Portuguese characters/BPE units. The models are trained for 150 epochs and each batch comprised of 7000 target units. The models form the last ten epochs are averaged and the averaged weight is used for decoding the hypothesis. The hypotheses are decoded with a beam of 10 and length penalty of 1. The beam search and EOS threshold described in section~\ref{section_Transformer_ASR} are used. The performance of the End-to-End SP-MT are tabulated in Table.~\ref{Transformer_End_to_End_SPMT}.

\begin{table}[!ht]
\renewcommand{\arraystretch}{1.5}
\setlength{\tabcolsep}{2pt}
\scriptsize
\begin{center}
\caption{Transformer based End-to-End SLT.}
\label{Transformer_End_to_End_SPMT}
\begin{tabular}{|c|c|c|}
\hline
Architecture~(Target unit)~(No.of parameters)&\multicolumn{1}{|c|}{Dev set}&Test set\\
\hline
S2S-Attention-BPE~\cite{visual_modality_for_MT}&-&36.0\\
\hline
Transformer-12enc-6dec-(char)(24M)&33.4&33.8\\
\hline
\shortstack{Transformer-12enc-8dec-(char)(29M)}&33.97&33.98\\
\hline
\shortstack{Transformer-8enc-6dec-(BPE)(24M}&36.91&37.45\\
\hline
\shortstack{Transformer-12enc-6dec-(BPE)(30M)}&38.95&39.06\\
\hline
\shortstack{Transformer-20enc-10dec-(BPE)(46M)}&40.03&40.19\\
\hline
\shortstack{Transformer-20enc-10dec-$d_{model}$-512}&39.94&40.59\\
\hline

\end{tabular}
\end{center}
\end{table}
From the Table.~\ref{Transformer_End_to_End_SPMT}, it can be observed that the models with BPE as target units has performed better than the models with characters as target units. The performance of these models is worse than the performance of the cascaded SLT models. To increase the model capacity, additional layers a have been added to encoder and decoder and the architecture of the model is in column~1. Row~8 of Table.~\ref{Transformer_End_to_End_SPMT} is a large transformer model with 20 encoder and 10 decoder layers and $d_{model}$ is 512 and rest of the models are trained $d_{model}$ is 256.\\

\subsection{Augmented Training for Cascaded SLT}
To reduce the mismatch between the ASR hypothesis and the oracle text, the training corpus for MT systems are augmented with the ASR hypothesis. ASR hypothesis for the training data is obtained using the ASR model described in section~\ref{section_Transformer_ASR}. The performance of these models is presented in Table~\ref{Augumented_ASR_MT_Transformer}.

\begin{table}[!ht]
\renewcommand{\arraystretch}{1.5}
\setlength{\tabcolsep}{2pt}
\scriptsize
\begin{center}
\caption{Cascaded SLT systems trained using augmented data.}
\label{Augumented_ASR_MT_Transformer}
\begin{tabular}{|c|c|c|c|c|}
\hline
~&\multicolumn{2}{|c|}{Oracle Text}&\multicolumn{2}{|c|}{ASR hypothesis}\\
\hline
~&Dev set&Test set&Dev set&Test set\\
\hline
Transformer-(char-char)&50.21&49.44&39.67&39.54\\
\hline
Transformer-(BPE-BPE)&49.59&49.41&39.12&39.82\\
\hline
Transformer-(char-BPE)&54.2&53.98&41.04&41.30\\
\hline
\end{tabular}
\end{center}
\end{table}
Rows~1,2 and 3 are the different cascaded SLT systems with different input/output granularites. Column 2, 3 are the performances of cascaded SLT system evaluated with oracle text and column 4, 5 are the performances of cascaded SLT systems evaluated with ASR hypothesis. From the table~.\ref{Augumented_ASR_MT_Transformer}, the performances of the systems trained with augmented training has not significantly improved the performance of the systems and has reduced the performance of the systems when evaluated with oracle text. The models trained from scratch with the augmented data and the models which are initially trained on clean data and later fine-tuned for augmented data have given the similar performance. 

\subsection{Using Averaged Embeddings}
Apart from relying on the individual embedding, if the model can also rely on the information from the context of the embedding then the model could be robust to noisy ASR hypothesis. To enforce this in the model, we have randomly selected 10\% of embedding from the input sequence and each embedding is averaged with the some other randomly selected embedding from the embedding matrix. This operation creates the uncertainty at the present embedding so that the model has to rely more on the context to retrieve information about the present label. This operation is done both at the encoder and decoder and the results are reported in the Table.\ref{Averaging_embedings_Transformer}.

\begin{table}[!ht]
\renewcommand{\arraystretch}{1.5}
\setlength{\tabcolsep}{2pt}
\scriptsize
\begin{center}
\caption{SLT models trained using Transformer by averaging the embeddings.}
\label{Averaging_embedings_Transformer}
\begin{tabular}{|c|c|c|c|c|c|c|c|c|}
\hline
~&\multicolumn{4}{|c|}{enc-random-avg}&\multicolumn{4}{|c|}{dec-random-avg}\\
\hline
~&\multicolumn{2}{|c|}{Oracle text}&\multicolumn{2}{|c|}{ASR hypothesis}&\multicolumn{2}{|c|}{Oracle text}&\multicolumn{2}{|c|}{ASR hypothesis}\\
\hline
~&Dev set&Test set&Dev set&Test set&Dev set&Test set&Dev set&Test set\\
\hline
char-char&51.94&50.61&39.12&39.18&53.44&52.02&40.1&39.47\\
\hline
BPE-BPE&50.63&50.2&40.46&40.63&52.44&51.57&41.31&41.23\\
\hline
char-BPE&53.8&52.6&41.27&41.25&54.87&55.79&41.85&41.38\\
\hline
\end{tabular}
\end{center}
\end{table}

In Table.\ref{Averaging_embedings_Transformer}, Column 2-5 are the performances of MT systems trained by averaging the encoder embeddings. Columns 2-3 and 4-5 are the performances of systems when evaluated using oracle text and ASR hypothesis text. Column 6-9 are the MT systems trained by averaging the decoder embeddings. The trained systems are evaluated using oracle text and ASR hypothesis and the results are presented in columns 6-7 and 8-9 respectively. From the Table.\ref{Averaging_embedings_Transformer}, it can be observed that the averaging the encoder embeddings has degraded the performance of the models. Averaging the embedings at the decoder has improved the performances of most of the models. Though there can be some improvements these improvements are not significantly better than the cascaded model.

\section{Multi-Task Training of SLT systems with ASR objective as an Auxiliary Loss}
\label{joint_traing}
\begin{figure}[!htbp]
	\begin{center}
	\includegraphics[width=50mm,height=50mm]{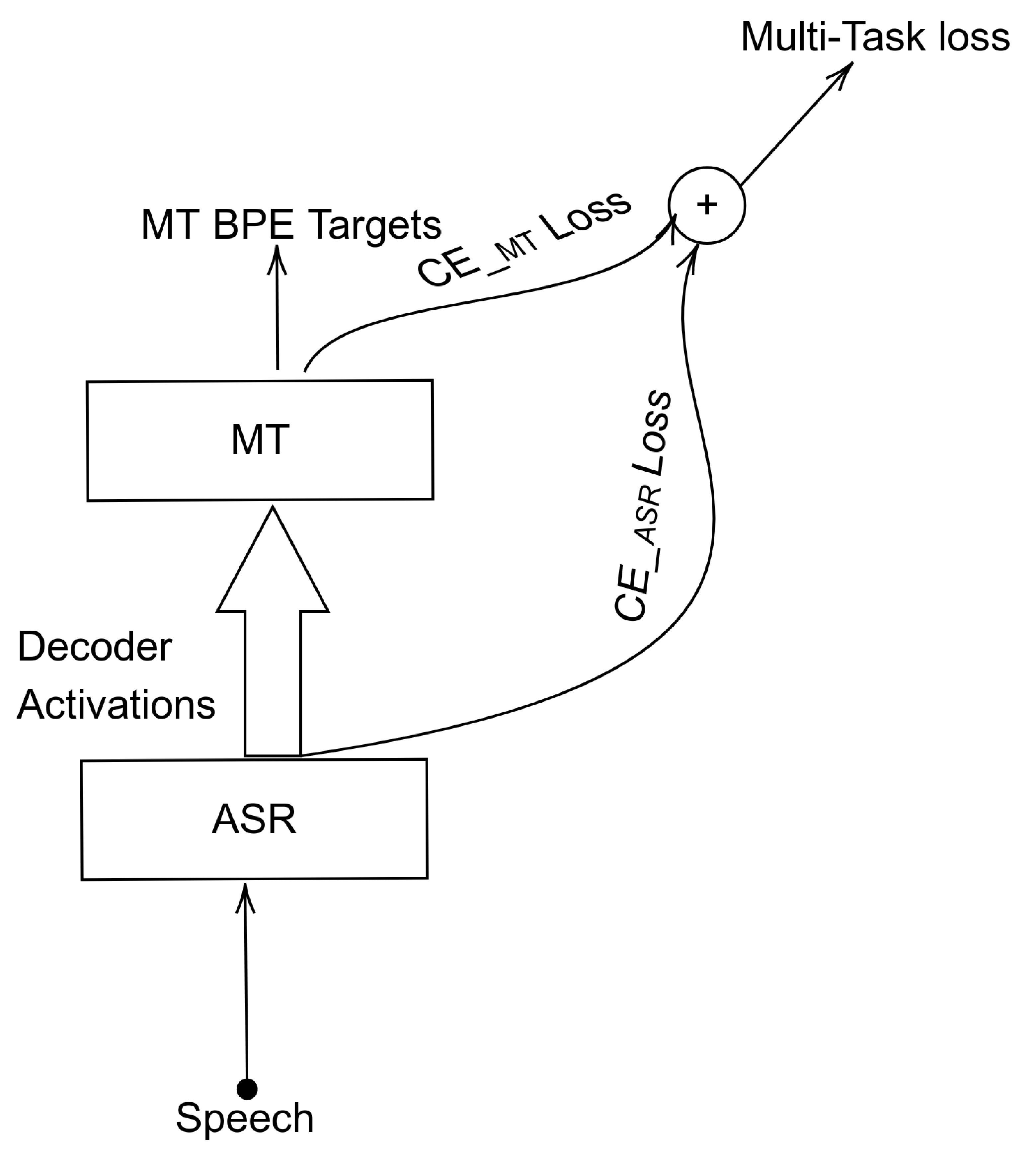}
	\caption{Block diagram of the Transformer based Sp-MT model with ASR objective as auxiliary loss function.}
	\label{multi_task_training}
	\end{center}
\end{figure}

From the above sections, it can be observed that the performance of the pipeline based systems is higher than the End-to-End approaches. Though the performance of the pipeline based approach is higher it can also observed that the performance gap of the MT systems when evaluated with oracle text and ASR hypothesis is higher than 10 BLEU score. To reduce this performance gap, we have trained a model with End-to-End differentiable pipeline between the spoken sequence and the target token sequence. This architecture uses the ASR objective as an auxiliary loss. The ASR and MT models described in the above sections are connected. The hidden representation from the ASR decoder is used as input to train the MT model. The block-diagram describing the model is presented in Figure.~\ref{multi_task_training}. This model would not make any discrete decisions from the ASR model, and the model is optimized for the final objective.

As shown in the block diagram, the neural hidden representation from the decoder of the ASR model is used as the input to the MT model. The models are optimized with a multi-task loss function. The ASR model in the pipeline has two gradients i.e., from ASR objective and MT objective and the MT model in the pipeline has gradients w.r.t to MT objective only. The models are trained for 150 epochs and the weights are averaged and averaged weights are used for decoding the model. The model has two decoders in the pipeline, the decoder produces the n-best hypothesis from the ASR with a beam size of 10 and neural hidden representation from each of the n-best output is considered as a separate input to the MT. The MT model produces the n-best hypothesis for each input with a beam size of 5. All the hypothesis produced by the MT model are combined and the best hypothesis from the MT model is used as the output hypothesis. The performance of the proposed SLT systems are tabulated in the Table.~\ref{Multi_task_training}. The block-diagram describing the proposed architecture is presented in Figure.\ref{Multi_task_training_with_ASR_auxilary_loss}.

\begin{table}[!ht]
\renewcommand{\arraystretch}{1.5}
\setlength{\tabcolsep}{2pt}
\scriptsize
\begin{center}
\caption{SLT systems jointly trained with a Multi-Task objective and ASR as an auxilary loss}
\label{Multi_task_training}
\begin{tabular}{|c|c|c|c|c|}
\hline
~&\multicolumn{2}{|c|}{Sacre-BLEU}&\multicolumn{2}{|c|}{WER}\\
\hline
~&Dev set&Test set&Dev set&Test set\\
\hline
Transformer-End-to-End&36.2&36.8&-&-\\
\hline
Transformer-(char-BPE)&40.48&40.46&24.27&24.50\\
\hline
Transformer-(BPE-BPE)&44.9&44.18&17.99&18.09\\
\hline
\end{tabular}
\end{center}
\end{table}

From the Table.~\ref{Multi_task_training}, it can be observed that the performance of this model is better than the standard pipeline systems. The proposed model has an improvement in the BLEU score of 4-5 compared to End-to-End systems. From the Table.\ref{Multi_task_training}, it can also be observed that the performance of the systems using both ASR and MT systems trained with BPE units is higher than the performance of the systems trained using ASR with characters and the MT systems with BPE units. This difference in the performance can be attributed to correspondence between the tokens of ASR and MT objectives. As the data-set has parallel text for training, there could be correspondence between the top 5k BPE units in English and Portuguese tokens and optimizing with the joint loss has optimized the ASR model with BPE as targets better than model optimized with characters as target tokens. From the Table~\ref{Multi_task_training}, the performance of the model is better than the pipeline based systems. This can also be observed from columns 3-4 of Table~\ref{Multi_task_training}, these are the WER's obtained from the jointly-trained model. From the joint training the ASR model with BPE target units has better performance than the ASR with character units and this has lead to better inference of the bottlenecks and the MT target sequence. To improve this performance the ASR and MT models could be ensembled with a separately trained ASR and MT models. The ensembling and joint decoding is presented in section~\ref{section_Multi_task_training_joint_decoding}.   

\section{SLT Systems with Multi-Task Training and Joint Decoding}
\label{section_Multi_task_training_joint_decoding}
\begin{figure}[!htbp]
	\begin{center}
		\includegraphics[width=50mm,height=50mm]{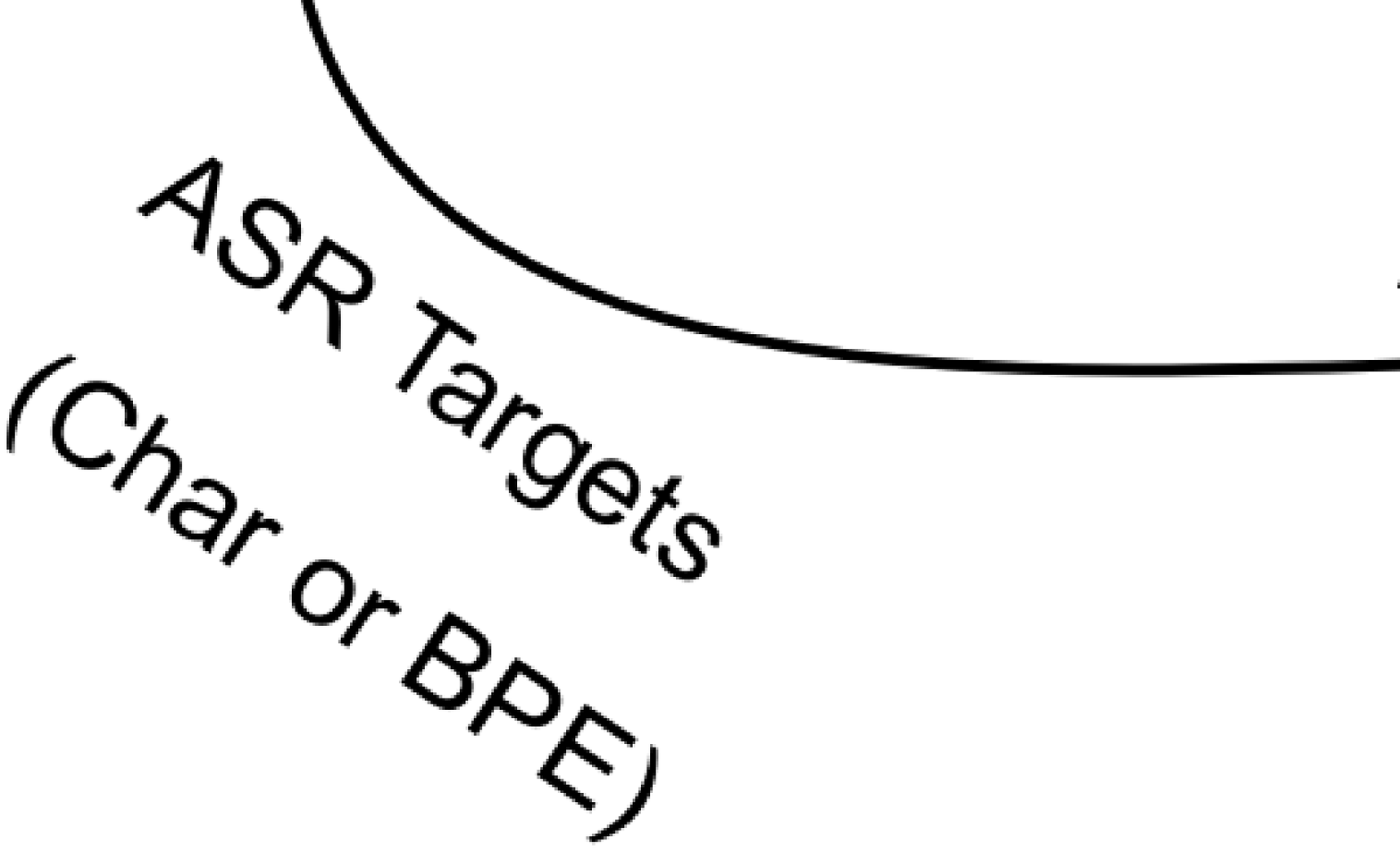}
		\caption{Block diagram describing the Multi-Task training with ASR objective as an auxiliary loss.}
		\label{Multi_task_training_with_ASR_auxilary_loss}
	\end{center}
\end{figure}

From Table.\ref{Multi_task_training}, it can be observed that the performance of the ASR model in the joint training is not on-par with the performance in subsection~\ref{section_Transformer_ASR}. To improve this performance the ASR model is ensembled with the ASR model trained in subsection~\ref{section_Transformer_ASR}. During the inference for the softmax distributions from both the models are computed for each prefix and the distributions are averaged and averaged distribution is used for the beam search. The proposed model also produces characters/BPE units as outputs along with the neural-hidden representations. Along the the neural hidden representations, the characters/BPE can also be used for translation and the MT models described in the subsection.~\ref{section_Transformer_machiene_translation} can also be is used to generate the translations. Both the models are ensembled and the performance are tabulated in the below Table.\ref{Multi_task_training_joint_decoding}. The Block diagram for the joint decoding is presented in Figure.~\ref{Multi_task_training_with_ASR_auxilary_loss}. ASR model is decoded with a beam size of 10 and every hypothesis is then decoded by a MT model with a beam size of~5.

\vspace{-1cm}

\begin{table}[!ht]
\renewcommand{\arraystretch}{1.5}
\setlength{\tabcolsep}{1pt}
\scriptsize
\begin{center}
\caption{Multi-Task training with ASR as a auxiliary loss and joint decoding}
\label{Multi_task_training_joint_decoding}
\begin{tabular}{|c|c|c||c|c||c|c||c|c|}
\hline
~&\multicolumn{2}{|c|}{stand-alone}&\multicolumn{2}{|c|}{Ens-ASR}&\multicolumn{2}{|c|}{Ens-MT}&\multicolumn{2}{|c|}{Ens-ASR-Ens-MT}\\
\hline
~&Dev set&Test set&Dev set&Test set&Dev set&Test set&Dev set&Test set\\
\hline
\shortstack{Transformer-\\(char-BPE)}&40.84&40.81&41.92&41.70&43.04&42.31&{\bf 43.72}&{\bf 43.06}\\
\hline
\shortstack{Transformer-\\(BPE-BPE)}&45.32&44.69&46.93&46.22&45.8&45.7&{\bf 47.33}&{\bf 46.9}\\
\hline
\end{tabular}
\end{center}
\end{table}
From Table~.\ref{Multi_task_training_joint_decoding},~column 2,~3 are the performances of jointly trained model. Column~4-5, 6-7 are the performances of SLT systems with ASR, MT ensembling respectively. Column 8-9 are the performances of SLT systems with both ASR and MT ensembling. From the table.~\ref{Multi_task_training_joint_decoding}, it can be observed that using joint decoding using both the models has improved the performances of both the models.

\section{Using Pre-trained ASR \& MT models for Multi-Task Training of SLT systems}
The SLT model described in section~\ref{joint_traing}, is initialized using the pre-trained ASR and MT models.The performance of these models is presented in Table.\ref{pretraining}. The pre-trained models are connected as described in the block diagram~\ref{multi_task_training}. The two models are connected with an additional linear layer or self-attention layers. The models are trained either by fine-tuning only the additional layers or fine-tuning the whole network. The models are trained for 100 epochs and decoded with same hyper-parameters as mentioned in section~\ref{joint_traing}.

\begin{table}[!ht]
\renewcommand{\arraystretch}{3}
\setlength{\tabcolsep}{2pt}
\scriptsize
\begin{center}
\caption{Multi-Task training with ASR as a auxilary loss. The models are initialized with pre-trained ASR and MT models.}
\label{pretraining}
\begin{tabular}{|c|c|c|c|c|c|c|c|c|}
\hline
~&\multicolumn{2}{|c|}{stand-alone}&\multicolumn{2}{|c|}{Ens-ASR}&\multicolumn{2}{|c|}{Ens-MT}&\multicolumn{2}{|c|}{\shortstack{Ens-ASR+\\Ens-MT}}\\
\hline
~&Dev&Test&Dev&Test&Dev&Test&Dev&Test\\
\hline
Transformer-(char-BPE)&40.84&40.81&41.92&41.70&43.04&42.31&\bf{43.72}&\bf{43.06}\\
\hline
\hline
\shortstack{Transformer-(char-BPE)\\pre-training+linear}&42.59&42.86&42.49&42.22&42.42&42.62&41.99&41.88\\
\hline
\shortstack{Transformer-(characters-BPE)\\pre-training+ self-attn layers}&42.74&42.87& 42.49&42.33&42.43&42.33&41.95&41.7\\
\hline
\shortstack{Transformer-(char-BPE)\\fine-tuning+linear}&43.18&42.91&43.81&43.22&43.08&42.67&43.36&42.98\\
\hline
\shortstack{Transformer-(char-BPE)\\fine-tuning+ self-attn layers}&42.65&42.93&42.82&42.82&{\bf 43.81}&{\bf43.22}&{\bf 43.36}&42.98\\
\hline
\hline
\shortstack{Transformer-(BPE-BPE)}&45.32&44.69&46.93&46.22&45.8&45.7&{\bf 47.33}&\bf{46.9}\\
\hline
\hline
\shortstack{Transformer-(BPE-BPE)\\pre-training+linear}&30.08&30.19&30.07&30.39&44.15&44.39&44.09&44.46\\
\hline
\shortstack{Transformer-(BPE-BPE)\\pretraining+selfattn}&45&44.7&45.15&44.99&44.67&44.84&44.73&44.79\\
\hline
\shortstack{Transformer-(BPE-BPE)\\finetuning+linear}&45.02&44.45&45.8&45.4&45.41&45.21&{\bf 46.12}&{\bf 45.97}\\
\hline
\shortstack{Transformer-(BPE-BPE)\\finetuning+selfattn}&45.05&44.7&45.35&45.77&45.23&45.05&45.6&45.5\\
\hline
\end{tabular}
\end{center}
\end{table}
From Table.\ref{pretraining}, row 3, 8 are the performances of systems which are trained from scratch. Rows 4,5  and 9,10 are the performances of the model obtained by fine-tuning the inserted layers. Rows 6,7 and 11, 12  are the performances obtained by fine-tuning whole pre-trained network along with the inserted layers. From Column~2, 3 of Table.~\ref{pretraining}, it can be observed the initializing the weights of the joint model from a separately trained ASR and MT has shown improvements, when the outputs of ASR are characters i.e., rows 4-7 have better performances than 3 but rows 9-11 have a slight degradation compared to row~8. This difference in performance can be attributed to the fact that the representations produced by the network for the characters are more generic to task compared to BPE units. This could also be seen in a sharp drop in the performance of stand-alone system in row~9 and the performance is recovered when the discrete BPE-units are decoded and translated through MT model. The joint models trained from scratch have shown better improvements when emsembled with other independent ASR and MT models as in rows~3,8. But the joint model with weights initialized from separately trained ASR and MT models as in rows ~4-7 and 9-12 did not have much improvements from ensembling as both the models are similar. Ensembling with MT models have shown better improvements in BLEU score compared to ensembling with ASR models. This improvements could be attributed to the reason that MT model in joint model uses the continuous representations from the ASR module while the ensembled MT models uses discrete representations, these modules are using different modalities and also the MT model is the last in the whole pipeline.  
Using the n-best hypothesis and doing a coupled search between ASR and MT has given better performances than using the 1-best hypothesis. Training the model in the proposed approach has give the best BLEU-score of 47.33 and 46.9  
on dev and test sets of HOW2 data-set. By far this is on-par with the best performing systems~\cite{KIT_IWSLT2019} in IWSLT-2019 on How2 data-set. The total parameters of the model (ASR+MT) are around 70M which is much lesser than the (>350M) parameters in the models~\cite{KIT_IWSLT2019}.

\section{Conclusion\& Future scope}
The performance gap between End-to-End SLT models have been improving but still pipeline based approaches are better than the End-to-End approaches. A large performance gap can be observed when machine translation models are evaluated with ASR hypothesis and oracle text transcription. Proposed systems aims to reduce this gap by training models with end-to-end differentiable pipeline between ASR and MT models. Using the ASR objective as auxiliary loss while optimizing the models have improved the performance of Joint SLT. The performance gains are higher when both ASR and MT models are trained with BPE as target units. As all the models are transformer models, they could be replaced with Non-Auto regressive models which could reduce the latency of decoding. A mechanism to force the transformer to retrieve the information from the context in the presence of less confident embedding could help the model to reduce the performance gap of MT models when used with oracle text and the noisy ASR hypothesis. While training with the noisy transcripts, using a sentence-wise confidence metric and lowering the scale of penalizing the model conditioning on the confidence metric could improve the performance of MT when dealing with noisy transcriptions. Using unpaired additional data for training ASR and MT models could improve the performance of the ensemble. 

\bibliographystyle{plain}
\bibliography{SP_MT.bib}
\end{document}